\documentclass[useAMS,usenatbib]{mn2e}
\usepackage[dvips]{epsfig}

\title[Probing the nature of USS sources]{Probing the nature of compact ultra-steep spectrum radio sources with the e-EVN and e-MERLIN}
\author[M.K. Argo et al.]{M.~K.~Argo$^{1}$, Z.~Paragi$^{2}$, H.~R\"ottgering$^{3}$, H.-R.~Kl\"ockner$^{4}$, G.~Miley$^{3}$, M.~Mahmud$^{5}$
\\
$^{1}$ASTRON, the Netherlands Institute for Radio Astronomy, Postbus 2, 7990 AA, Dwingeloo, The Netherlands\\ 
$^{2}$Joint Institute for VLBI in Europe (JIVE), Postbus 2, 7990 AA, Dwingeloo, The Netherlands\\
$^{3}$Leiden Observatory, Universiteit Leiden, Niels Bohrweg 2, NL-2333 CA Leiden, The Netherlands\\
$^{4}$Max-Planck-Institute f\"ur Radioastronomie, Auf dem H\"ugel 69, 53121 Bonn, Germany\\
$^{5}$Formerly at JIVE\\
}

\begin{document}

\date{Revision January 2013}
\pagerange{\pageref{firstpage}--\pageref{lastpage}} \pubyear{2012}
\maketitle
\label{firstpage}

\begin{abstract}
We present first results from electronic Multi-Element Remotely Linked Interferometer Network (e-MERLIN) and electronic European VLBI Network (e-EVN) observations of a small sample of ultra-steep spectrum radio sources, defined as those sources with a spectral index $\alpha < -1.4$ between 74\,MHz and 325\,MHz, which are unresolved on arcsecond scales.  Such sources are currently poorly understood and a number of theories as to their origin have been proposed in the literature. The new observations described here have resulted in the first detection of two of these sources at milliarcsecond scales and show that a significant fraction of ultra-steep spectrum sources may have compact structures which can only be studied at the high resolution available with very long baseline interferometry (VLBI).
\end{abstract}


\section{Introduction}
Radio sources with an ultra steep radio spectrum (USS; defined here as having $\alpha < -1.4$ where S$_{\nu} \propto \nu^{\alpha}$) are often associated with extreme phenomena in the universe.  USS sources with a classical FRII type radio morphology (highly collimated jets and well-defined lobes with prominent hotspots) are, for the most part, associated with very distant galaxies ($2 < z < 5.2$; \citealt{breuck00,rott97}).  Being very massive and often located in proto-clusters, these sources can be used to study the origin and evolution of massive galaxies and clusters.  Recent high-redshift results include (i) the finding that massive radio 
galaxies with masses of around $10^{12}$\,M$_{\odot}$ exist at up to $z \sim 5$ \citep{seymour07}, and (ii) the discovery of proto-clusters up to $z \sim 4$ with total masses as expected for progenitors of nearby clusters (a few $10^{14}$\,M$_{\odot}$; \citealt{venemans07}).  USS sources with mostly diffuse radio emission are virtually always associated with close-by clusters up to $z \sim 0.5$ \citep{weeren09} and the emitting regions can have spectacular sizes up of to a few Mpc.  The most likely explanation for this class of object is radio emitting plasma tracing shocks in merging clusters.  The study of these two types of USS source started in the late seventies with studies of their morphology, and follow-up work with large optical and X-ray telescopes subsequently gave an understanding of their enormous importance for studies of galaxy and cluster evolution.  

A third class of USS has not yet received much attention: those which are unresolved on arcsecond scales.  The aim of the observations 
presented here is to take a first step towards determining the nature of these objects. There are a great number of possible scenarios for the origin of such compact USS objects including (i) radio galaxies located at or near the epoch of re-ionisation (it has been suggested that objects with the steepest spectral index correspond to the highest redshift objects, e.g. \citealt{krolik91}), (ii) young obscured radio galaxies with the early phase perhaps accompanied by a starburst (\citealt{reuland04} inferred star formation rates of $\sim$1000\,M$_{\odot}$/year), (iii) steep spectrum core quasars with the steepening perhaps caused by cooling of the plasma via scattering of infrared photons from the torus by the synchrotron-emitting electrons, and (iv) pulsars (\citealt{breuck00} found a clear over density of sources with $\alpha<-1.6$ near the Galactic plane).
Long baseline radio interferometric observations may give very important clues about the nature of these objects. Star formation activity in galaxies is expected to produce low brightness temperature emission ($\sim10^4$~K, see \citealt{alexandroff12} and references therein) and diffuse morphology.  Synchrotron emission from radio lobes on kiloparsec scales can be very well imaged with e-MERLIN, but will still be mostly resolved out on milliarcsecond scales. Finally, steep spectrum quasars and pulsars will be compact when probed with VLBI; the potential Galactic origin of the source can then be verified by multi-epoch proper motion measurements.
This letter describes e-MERLIN and e-EVN observations of a sample of compact ultra-steep spectrum sources, aimed at attempting to investigate their morphologies and narrow down the possibilities.  Section \ref{SecObs} describes the sample selection and observations, the results are discussed in section \ref{SecResults} and in section \ref{SecDiscussion} we draw some conclusions.

{\small
\centering
\begin{table*} 
\label{USSsample}
\caption{Compact sources with a spectral index $\alpha^{74\,\rm MHz}_{325 \, \rm  MHz} < -1.4 $ observed with the e-EVN.
Names correspond to the NVSS catalogue names.  The coordinates are from the VLA FIRST Survey. The spectral index is calculated from the VLSS and WENSS flux densities, and the flux density ratio is calculated from the NVSS and FIRST catalogues.  The VLSS, WENSS, NVSS and FIRST flux densities are given in mJy. Positions are given in J2000 coordinates. Note that both the WENSS flux density and spectral index of J151229+470245 were revised after the observations due to an error discovered in one of the catalogues (see text).}
\begin{tabular}{ccccccccc}
\hline
\hline
Name & ----- RA ----- & ----- Dec ----- & $\alpha^{\rm 74\,MHz}_{\rm 325\,MHz}$  & ${S^{\rm NVSS}_{\rm 1400\,MHz}} \over {\rm S}^{\rm FIRST}_{\rm 1400\,MHz}$&$S^{\rm VLSS}_{\rm 74\,MHz}$ & $S^{\rm WENSS}_{\rm 325\,MHz}$ & $S^{\rm NVSS}_{\rm 1400\,MHz}$ & $S^{\rm FIRST}_{\rm 1400\,MHz}$\\
\hline
J072212+291042 & 07:22:12.603 & +29:10:41.51 &  $-$1.45 &    1.1 &  650 &   76 &   11 &   10\\
J082916+383453 & 08:29:17.351 & +38:34:52.62 &  $-$1.44 &    0.9 &  940 &  112 &   11 &   12\\
J130612+514407 & 13:06:12.176 & +51:44:06.93 &  $-$1.58 &    1.1 & 1640 &  159 &   27 &   25\\
J131655+483200 & 13:16:55.642 & +48:32:00.13 &  $-$1.52 &    1.1 & 2390 &  252 &   67 &   61\\
J151229+470245 & 15:12:29.174 & +47:02:45.59 &  $-$0.50 &    1.0 & 1130 &  543 &  217 &  221\\
\hline
\hline
\end{tabular} 
\end{table*} 
}

\section{Sample selection and observations}\label{SecObs}

The initial sample of USS sources was found by correlating the 74\,MHz VLSS catalogue \citep{cohen07} against the 325\,MHz WENSS database \citep{rengelink97} and selecting the small fraction of VLSS sources with a sufficiently steep spectrum. A sub-sample of compact sources was then selected by choosing objects which were both unresolved in the FIRST survey \citep{becker95}, and had a flux ratio between the NVSS \citep{condon98} and FIRST catalogues of $0.8 < R < 1.2$, ensuring that 
the 1.4\,GHz emission is indeed compact on scales of $\sim$1\,arcsecond.  The FIRST and NVSS maps were also visually inspected to ensure that the sources were unresolved and that there were no nearby objects which could have led to ambiguities in the measured fluxes. The resulting sample contained 17 sources. A sub-sample of five sources was selected for an exploratory VLBI survey, all of these sources had NVSS flux densities larger than 10\,mJy (see Table~\ref{USSsample}).

\subsection{e-EVN and WSRT observations}

The five sources (J072212+291042, J082916+383453, J130612+514407, J131655+483200 and J151229+470245) were observed with the EVN at 1.6~GHz in e-VLBI mode \citep{szomoru08} over two 10-hour runs carried out on 10$^{\rm th}$ and 11$^{\rm th}$ June 2010 (programmes EP070A and B; PI R\"ottgering).  Ten stations participated in the experiment: Effelsberg, Medicina, Onsala, Torun, Westerbork (12 antennas tied array), Lovell, Cambridge, Darnhall, Knockin and Sheshan. The target sources were phase-referenced to nearby calibrators;
the typical on-source time was 2.5 hours. The total aggregate bitrate per telescope was 1024\,Mb/s, except for the MERLIN antennas which participated with a lower rate of 128~Mbps. Both LCP and RCP polarizations were observed with 2-bit sampling. The data were processed using standard procedures following the EVN Data Analysis 
Guide\footnote{http:/$\!$/www.evlbi.org/user\_guide/guide/userguide.html}.

During the e-EVN observations, synthesis array data were also recorded at Westerbork and were independently processed in AIPS.  The data were recorded using 8 bands of 20\,MHz, each split into 64 channels.  For amplitude calibration we used 3C286. There was significant interference during the observation resulting in the loss of the first two subbands.

Out of the five objects from the sample which were observed with the e-EVN, J072212+291042, J130612+514407 and J151229+470245 were detected.  Note that initially J130612+514407 was not detected because of a 12.8-arcsecond error in the pointing position; a clear detection of the source was obtained only after this error was discovered and a wide-field image was made.  Table \ref{detections} summarises the detections of these sources.  Since, for the sources which were detected, the e-EVN observations 
recovered only $\sim$20\% of the VLA flux at the same frequency, e-MERLIN commissioning observations were requested in order to investigate their structures on intermediate scales.  Comments on each of these sources are given in the next section.

\subsection{e-MERLIN observations}

e-MERLIN observations of two of the EVN-detected sources were requested in order to investigate their structures on intermediate scales.  The e-MERLIN observations were carried out in March 2011, during the commissioning period. At this point in the commissioning phase, five stations of the array were available with the new C-band (4- to 6-GHz) system: Mk2, Tabley, Darnhall, Knockin and Defford.  Observations were carried out at 6.6\,GHz using the new capabilities of the upgraded e-MERLIN array\footnote{See http://www.e-merlin.ac.uk/tech/ for the current capabilities of the e-MERLIN array.} and used four adjacent bands of 128\,MHz, each split into 512 channels, starting at 6.64\,GHz with a central frequency of 6.89\,GHz.  In both cases 3C286 was used as the primary flux calibrator and OQ208 was used as the bandpass calibrator.  Both sources were observed for $\sim$4.5 hours on 26th March and $\sim$10 hours on March 30th 2011. J151229+470245 used the source 1500+478 for phase 
referencing,  J072212+291042 used 0736+299 as the phase calibrator.  Of the total observing time, only $\sim$9 hours on each source were ultimately usable due to a variety of commissioning issues.  The data were flagged spectrally, averaged in frequency, fringe fitted and then calibrated using standard methods for phase referencing experiments.

\section{Results}\label{SecResults}

\subsection{J072212+291042}

With a flux density of 10\,mJy in FIRST and 11\,mJy in the NVSS (Table \ref{detections}), J072212+291042 is the weakest of the three VLBI detections.  The e-EVN observations showed a $\sim$2\,mJy source with no clear evidence of structure on VLBI scales, consistent with a point source model, albeit at significantly lower flux density than that detected by the VLA and Westerbork: only $\sim$30\% of the Westerbork flux is recovered by the e-EVN at the same frequency indicating the presence of significant structure on intermediate scales.

The source is not detected with e-MERLIN at 6.9\,GHz to a 3$\sigma$ limit of 0.12\,mJy/beam.  Comparing the 3-$\sigma$ limit at 6.9\,GHz with the flux detected by the e-EVN at 1.4\,GHz gives a limit on the spectral index of $\alpha < -1.4$, showing that, in this case, the spectral index remains steep at higher frequencies.  A search of the catalogues in the NASA/IPAC Extragalactic Database (NED)\footnote{http://ned.ipac.caltech.edu/} shows that J072212+291042 has no known counterparts in other wavebands.
The compact nature of this object suggests that the most likely explanation is either a steep spectrum quasar core or a Galactic pulsar.  Although this source has a low galactic latitude (see Table \ref{detections}) a search of pulsar catalogues around this location shows no currently known pulsar or pulsar candidate at these coordinates (Eatough, private communication).

\subsection{J130612+514407}

Of the detected sources, J130612+514407 has the steepest spectral index at low frequencies.  Without an observation at 6.8\,GHz, however, it is impossible to tell if the spectrum remains steep at higher frequencies - the positional offset in the e-EVN observations was only discovered after the e-MERLIN observations had taken place.  The e-EVN data show an unresolved point source with an integrated flux density of 7.1\,mJy, recovering $\sim30\%$ of the emission detected by the Westerbork array during the same observation, implying structure on intermediate scales at 1.4\,GHz.
However, the primary beam of the Westerbork tied-array is significantly smaller than the 12.8" offset of the source position from the pointing centre of the e-EVN observation; imaging the source without including the Westerbork array results in a peak flux density of 18.5 mJy and an integrated intensity of 21.6 mJy, recovering almost all of the WSRT-only flux, although it should be noted that these measurements are somewhat uncertain due to the large positional offset.

Unlike J072212+291042, J130612+514407 has an optical counterpart in SDSS, J130612.15+514407.0, a diffuse-looking galaxy with an SDSS spectrum giving a redshift of 0.2773.  The source also has a counterpart in the JVAS/CLASS sample with an integrated flux density at 8.4\,GHz of 16.2\,mJy.  If the source does not vary, then this results in a spectral index of $-$0.31 between 1.4 (NVSS) and 8.4\,GHz making it slightly steep, but certainly not ultra-steep at these frequencies.  However, the CLASS observations were carried out with the VLA in its largest A configuration resulting in an angular resolution of 220 milli-arcseconds (mas), very different to the beam size of the NVSS survey, making this spectral index measurement unreliable.  This object also has counterparts in 2MASS and the ROSAT X-ray catalogue, and is classified as the brightest cluster galaxy in \cite{koester07}, making the quasar core scenario the most likely.

\subsection{J151229+470245}

\begin{figure}
\centering
\includegraphics[width=8cm,angle=270]{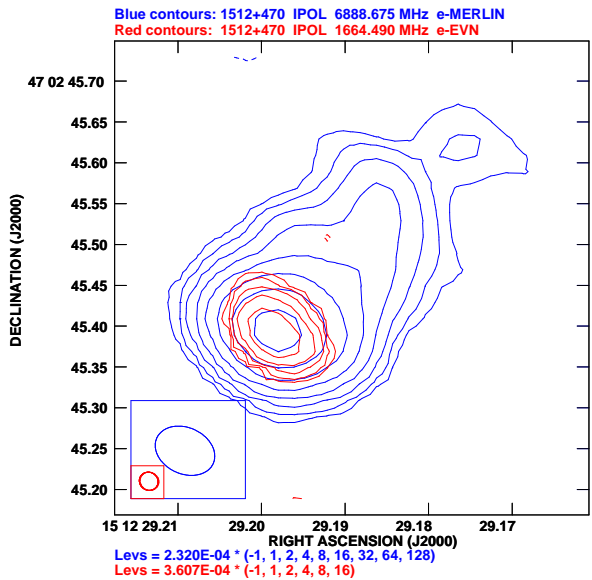}
\caption{\label{fig1512}e-MERLIN map of J151229+470245 at 6.9\,GHz (blue contours) together with the e-EVN map at 1.4\,GHz (red contours).  
Contours are plotted at -1,1,2,4,8,16,32,64,128 $\times$ 0.23\,mJy/beam for the e-MERLIN data and -1,1,2,4,8,16 $\times$ 0.36\,mJy/beam for the e-EVN data.  The boxes in the lower left indicate the size of the restoring beam for each image.}
\end{figure}

In contrast, J151229+470245 is clearly resolved in both the e-EVN and e-MERLIN images (see Fig. \ref{fig1512}). The WSRT-only data show an unresolved source, as expected from the NVSS/FIRST images, with a peak flux of 192\,mJy/beam and an integrated flux density of 203\,mJy.  This is within 8\% of the flux recorded in the FIRST and NVSS catalogues.  The e-EVN observation shows a source elongated in a NE-SW direction, with a peak brightness to total flux density ratio of 1:4. 

The map obtained with e-MERLIN at 6.9\,GHz (Fig. \ref{fig1512}) shows a very different structure.  The source is clearly resolved and appears to show a core-jet morphology.  The e-EVN component is coincident with the brightest feature observed with e-MERLIN.  A two-component fit was carried out on the e-MERLIN map using the 
AIPS task {\sc jmfit}: a relatively compact (compared to the e-MERLIN beam), bright Gaussian with a peak of 39.5\,mJy/beam, an integrated flux of 66.1\,mJy and a size of 98$\times$73 mas at a position angle of 71 degrees, and a more extended Gaussian ``jet" component with a fitted peak brightness of 4.14\,mJy/beam, an integrated flux density of 28.3\,mJy with a size of 0.2$\times$0.12 arcseconds at a position angle of 145 degrees.

The nature of this source is rather puzzling.  Interestingly, while the e-MERLIN map (taken on its own) appears to show a classic core-jet morphology, the e-EVN (``core") component is very resolved suggesting that it is not a simple AGN core. Unless there is very strong scatter broadening of the target, the observed structure is more consistent with lobe emission detected on 10-mas scales. The observed steep spectrum, low brightness temperature and no variability (between VLA and WSRT epochs) all support this scenario. Therefore we may have identified a ``naked jet-lobe'' source without strong AGN core emission.  The absence of a compact, self-absorbed core is a sign of a Type 2 AGN, when the jet is viewed at a large angle compared to our line of sight.
It is intriguing though, why such an unbeamed object would have a prominent one-sided jet morphology.  An alternative interpretation is that we are observing a two-sided jet-lobe with a very weak core within the structure.

On inspection of the WENSS map, we found that J151229+470245 is a complex source at the resolution of Westerbork at 325\,MHz.  An examination of the map in comparison with the survey database, the corresponding fields in both NVSS and FIRST, and the new Westerbork data obtained during the e-EVN observations, determined that the fluxes listed in the WENSS catalogue do not refer to the source components at the listed coordinates.  The calibrated fits file for the WENSS field was retrieved and Gaussian components were fitted to the three sources in the region.  This showed that the source in which we are interested actually had an integrated flux of 543\,mJy, not 123\,mJy as listed in the WENSS catalogue. 
This gives the source a spectral index of -0.50 between 74 and 325\,MHz, making it steep spectrum, but not ultra-steep.  The spectrum is slightly steeper between WENSS and FIRST/NVSS with $\alpha$ = -0.62 between 325\,MHz and 1.4\,GHz. Therefore, based on its spectrum and the observed VLBI and e-MERLIN radio structure we may classify J151229+470245 as a peculiar compact steep spectrum radio source (CSS).

This source is also an example of the class of radio sources known as Infrared Faint Radio Sources (IFRS), those with clear radio emission but no corresponding infrared detection \cite[e.g.][]{norris06}.  Such sources are difficult to identify due to their lack of counterparts outside the radio regime and, as yet, no IFRS source has a measured redshift, but they are thought to be high redshift quasars due to their generally compact nature, mostly steep spectral indexes and high brightness temperatures.  While some of the known IFRS sources are extended on arcsecond scales with the VLA, and a small number of such sources have previously been detected at VLBI-resolution (e.g. \citealt{norris07,middelberg08}), this appears to be the only example so far which is resolved on VLBI scales.

{\small
\centering
\begin{table*} 
\caption{\label{detections}e-EVN detected sources. The measured coordinates in the VLBI experiment ($\pm0.1$\,mas) and the corresponding galactic coordinates are both given.  Flux densities are given for each observation: WSRT (1.6 GHz), e-EVN (1.6 GHz) and e-MERLIN (6.8 GHz), and the NVSS-WSRT spectral index, and VLBI-WSRT flux density ratios are calculated.  For J072212+291042 a 3$\sigma$ upper limit is given at 6.89\,GHz.  Note that the WSRT and VLBI fluxes at 1.6\,GHz were obtained simultaneously, and that J130612+514407 was not observed with e-MERLIN.}
\begin{tabular}{ccccccccc}
\hline
\hline
Name & RA  & Dec & ($l$,$b$) &  $S^{\rm WSRT}_{1670 \rm MHz}$ & $S^{\rm{e\textrm{-}EVN}}_{1670 \rm MHz}$ &
$S^{\rm{e\textrm{-}MERLIN}}_{6890 \rm MHz}$ & $\alpha^{1400\,\rm MHz}_{1670 \, \rm  MHz}$ & $S^{\rm{e\textrm{-}EVN}}_{1670 \rm MHz} \over S^{\rm WSRT}_{1670 \rm MHz}$\\
\hline
J072212+291042 & 07:22:12.60906 & +29:10:41.6648 & 189, 19 &   7.6 &   2.1 &  $<$0.12 &  $-1.82$ &  0.28\\
J130612+514407 & 13:06:12.16262 & +51:44:06.9627 & 118, 65 &  23.1 &  21.6 &      $-$ &  $-1.28$ &  0.94\\
J151229+470245 & 15:12:29.19905 & +47:02:45.3837 &  78, 56 & 203   &  42.0 &     94.0 &  $-0.51$ &  0.21\\
\hline
\hline
\end{tabular} 
\end{table*} 
}

\section{Discussion and conclusions}\label{SecDiscussion}

Of the five sources in the sample, one, J151229+470245, is not an ultra steep spectrum source.  Of the remaining four, two are undetected with VLBI, two are detected but are unresolved on milli-arcsecond scales.  These new high resolution data are not sufficient to uniquely determine the nature of these sources, however they do provide valuable information on the variety of sources one may find in a larger USS sample.  As one may expect, a significant fraction of these sources are completely resolved out since steep spectrum emission is often related to very extended lobes in radio galaxies.  A detection rate of fifty per cent, however, indicates that very high resolution studies of a moderate fraction of USS sources will be possible. 

Both of our USS source detections, J072212+291042 and J130612+514407, are consistent with a point source on VLBI scales. Note that in these short, exploratory observations with limited $uv$-coverage (especially on the long baselines) we cannot adequately constrain the source sizes, but they do appear more compact than our beam of 32$\times$26 mas. The emission in these two cases is, therefore, consistent with either a Galactic pulsar or a very steep spectrum quasar core. The former hypothesis will be easily tested by further VLBI observations that can potentially show the proper motion of the pulsar in our Galaxy.

Since there are generally no detections of these sources in any other wavelength regime and no redshift information is available, they are very difficult to classify.  These new radio observations provide fresh information and vital clues in the search for an understanding of their nature.

J072212+291042 is compact on VLBI scales but the e-EVN observations presented here recover less than 20\% of the flux seen at the same frequency on VLA scales, clearly indicating the existence of structure on intermediate spatial scales to which neither of these arrays are sensitive.  e-MERLIN observations at 6.9\,GHz fail to detect the source to a 3-$\sigma$ limit of 120\,$\mu$Jy/beam, confirming the steep spectrum extends to higher frequencies.  Observations with e-MERLIN at 1.4\,GHz would probe this intermediate-scale emission, but this band was not available at the time when the commissioning observations described here were obtained.
The fact that only $\sim$20 per cent of the FIRST/NVSS flux is recovered on VLBI scales strongly suggests the existence of structure on intermediate scales for these sources.  A study by \cite{kloeckner09} of a sample of 11 high-redshift obscured quasars also found similar evidence for structures on intermediate scales, although the VLBI-recovered fluxes there ranged from 30 to 100 per cent of the low-resolution measurements.

J130612+514407 is also compact on VLBI scales with significantly less flux lost when imaged with long baselines.  Unlike J072212+291042, however, this source does have counterparts in other wavebands which suggest the source is associated with a galaxy at a redshift of 0.2773.

While J151229+470245 appears to be a compact steep spectrum object, rather than an ultra steep spectrum source, it is still compact with a steep spectrum at low frequencies and interesting structure at higher resolutions.  VLBI observations of J151229+470245 only recover 20\% of the NVSS flux.  In sharp contrast to J072212+291042 and J130612+514407 however, the object is clearly resolved on both e-MERLIN and VLBI scales, showing different structures in the two maps.  While the e-MERLIN map shows what appears to be a classic core-jet morphology, the ``core" is clearly resolved in the e-EVN map.  Observations with e-MERLIN at L-band (1.4\,GHz) would again allow us to probe the morphology of the missing flux.

While firm conclusions regarding the nature of these sources remain elusive, the observations and unexpected results presented here do show that compact ultra-steep spectrum sources are a diverse class of object whose nature can be probed with VLBI.

\section*{Acknowledgments}

The authors thank Ralph Eatough for useful discussions.
e-MERLIN is a National Facility operated by the University of Manchester at Jodrell Bank Observatory on behalf of STFC.
We thank the e-MERLIN operations team for data taken during commissioning time. 
The EVN is a joint facility of European, Chinese, South African and other radio astronomy institutes funded by their national research councils. 
e-VLBI research infrastructure in Europe is supported by the European Union's Seventh Framework Programme (FP7/2007-2013) under grant agreement RI-261525 NEXPReS.
The WSRT is operated by ASTRON (Netherlands Institute for Radio Astronomy) with support from the Netherlands foundation for Scientific Research.
This research has made use of the VizieR catalogue access tool, CDS, Strasbourg, France. The original description of the VizieR service was published 
in A\&AS 143, 23.
This research also made use of the NASA/IPAC Extragalactic Database (NED) which is operated by the Jet Propulsion Laboratory, California Institute of Technology, 
under contract with the National Aeronautics and Space Administration.

\bibliographystyle{mn}
\bibliography{refs}

\begin{thebibliography}{18}
\expandafter\ifx\csname natexlab\endcsname\relax\def\natexlab#1{#1}\fi

\bibitem[{{Alexandroff} {et~al.}(2012){Alexandroff}, {Overzier}, {Paragi},
  {Basu-Zych}, {Heckman}, {Kauffmann}, {Bourke}, {Lobanov}, {Ptak}, \&
  {Schiminovich}}]{alexandroff12}
{Alexandroff} R., {Overzier} R.~A., {Paragi} Z., {Basu-Zych} A., {Heckman} T.,
  {Kauffmann} G., {Bourke} S., {Lobanov} A., {Ptak} A., {Schiminovich} D.,
  2012, {Mon. Not. R. Astr. Soc.}, 423, 1325

\bibitem[{{Becker} {et~al.}(1995){Becker}, {White}, \& {Helfand}}]{becker95}
{Becker} R.~H., {White} R.~L., {Helfand} D.~J., 1995, {Astrophys. J.}, 450, 559

\bibitem[{{Cohen} {et~al.}(2007){Cohen}, {Lane}, {Cotton}, {Kassim}, {Lazio},
  {Perley}, {Condon}, \& {Erickson}}]{cohen07}
{Cohen} A.~S., {Lane} W.~M., {Cotton} W.~D., {Kassim} N.~E., {Lazio} T.~J.~W.,
  {Perley} R.~A., {Condon} J.~J., {Erickson} W.~C., 2007, {Astron. J.}, 134,
  1245

\bibitem[{{Condon} {et~al.}(1998){Condon}, {Cotton}, {Greisen}, {Yin},
  {Perley}, {Taylor}, \& {Broderick}}]{condon98}
{Condon} J.~J., {Cotton} W.~D., {Greisen} E.~W., {Yin} Q.~F., {Perley} R.~A.,
  {Taylor} G.~B., {Broderick} J.~J., 1998, {Astron. J.}, 115, 1693

\bibitem[{{De Breuck} {et~al.}(2000){De Breuck}, {van Breugel},
  {R{\"o}ttgering}, \& {Miley}}]{breuck00}
{De Breuck} C., {van Breugel} W., {R{\"o}ttgering} H.~J.~A., {Miley} G., 2000,
  {Astronomy and Astrophysics Supplement Series}, 143, 303

\bibitem[{{Kl{\"o}ckner} {et~al.}(2009){Kl{\"o}ckner},
  {Mart{\'{\i}}nez-Sansigre}, {Rawlings}, \& {Garrett}}]{kloeckner09}
{Kl{\"o}ckner} H.-R., {Mart{\'{\i}}nez-Sansigre} A., {Rawlings} S., {Garrett}
  M.~A., 2009, {Mon. Not. R. Astr. Soc.}, 398, 176

\bibitem[{{Koester} {et~al.}(2007){Koester}, {McKay}, {Annis}, {Wechsler},
  {Evrard}, {Bleem}, {Becker}, {Johnston}, {Sheldon}, {Nichol}, {Miller},
  {Scranton}, {Bahcall}, {Barentine}, {Brewington}, {Brinkmann}, {Harvanek},
  {Kleinman}, {Krzesinski}, {Long}, {Nitta}, {Schneider}, {Sneddin}, {Voges},
  \& {York}}]{koester07}
{Koester} B.~P., {McKay} T.~A., {Annis} J., {Wechsler} R.~H., {Evrard} A.,
  {Bleem} L., {Becker} M., {Johnston} D., {Sheldon} E., {Nichol} R., {Miller}
  C., {Scranton} R., {Bahcall} N., {Barentine} J., {Brewington} H., {Brinkmann}
  J., {Harvanek} M., {Kleinman} S., {Krzesinski} J., {Long} D., {Nitta} A.,
  {Schneider} D.~P., {Sneddin} S., {Voges} W., {York} D., 2007, {Astrophys.
  J.}, 660, 239

\bibitem[{{Krolik} \& {Chen}(1991)}]{krolik91}
{Krolik} J.~H., {Chen} W., 1991, {Astron. J.}, 102, 1659

\bibitem[{{Middelberg} {et~al.}(2008){Middelberg}, {Norris}, {Tingay}, {Mao},
  {Phillips}, \& {Hotan}}]{middelberg08}
{Middelberg} E., {Norris} R.~P., {Tingay} S., {Mao} M.~Y., {Phillips} C.~J.,
  {Hotan} A.~W., 2008, {Astron. Astrophys.}, 491, 435

\bibitem[{{Norris} {et~al.}(2006){Norris}, {Afonso}, {Appleton}, {Boyle},
  {Ciliegi}, {Croom}, {Huynh}, {Jackson}, {Koekemoer}, {Lonsdale},
  {Middelberg}, {Mobasher}, {Oliver}, {Polletta}, {Siana}, {Smail}, \&
  {Voronkov}}]{norris06}
{Norris} R.~P., {Afonso} J., {Appleton} P.~N., {Boyle} B.~J., {Ciliegi} P.,
  {Croom} S.~M., {Huynh} M.~T., {Jackson} C.~A., {Koekemoer} A.~M., {Lonsdale}
  C.~J., {Middelberg} E., {Mobasher} B., {Oliver} S.~J., {Polletta} M., {Siana}
  B.~D., {Smail} I., {Voronkov} M.~A., 2006, {Astron. J.}, 132, 2409

\bibitem[{{Norris} {et~al.}(2007){Norris}, {Tingay}, {Phillips}, {Middelberg},
  {Deller}, \& {Appleton}}]{norris07}
{Norris} R.~P., {Tingay} S., {Phillips} C., {Middelberg} E., {Deller} A.,
  {Appleton} P.~N., 2007, {Mon. Not. R. Astr. Soc.}, 378, 1434

\bibitem[{{Rengelink} {et~al.}(1997){Rengelink}, {Tang}, {de Bruyn}, {Miley},
  {Bremer}, {R\"ottgering}, \& {Bremer}}]{rengelink97}
{Rengelink} R.~B., {Tang} Y., {de Bruyn} A.~G., {Miley} G.~K., {Bremer} M.~N.,
  {R\"ottgering} H.~J.~A., {Bremer} M.~A.~R., 1997, {Astronomy and Astrophysics
  Supplement Series}, 124, 259

\bibitem[{{Reuland} {et~al.}(2004){Reuland}, {R{\"o}ttgering}, {van Breugel},
  \& {De Breuck}}]{reuland04}
{Reuland} M., {R{\"o}ttgering} H., {van Breugel} W., {De Breuck} C., 2004,
  {Mon. Not. R. Astr. Soc.}, 353, 377

\bibitem[{{R\"ottgering} {et~al.}(1997){R\"ottgering}, {van Ojik}, {Miley},
  {Chambers}, {van Breugel}, \& {de Koff}}]{rott97}
{R\"ottgering} H.~J.~A., {van Ojik} R., {Miley} G.~K., {Chambers} K.~C., {van
  Breugel} W.~J.~M., {de Koff} S., 1997, {Astron. Astrophys.}, 326, 505

\bibitem[{{Seymour} {et~al.}(2007){Seymour}, {Stern}, {De Breuck}, {Vernet},
  {Rettura}, {Dickinson}, {Dey}, {Eisenhardt}, {Fosbury}, {Lacy}, {McCarthy},
  {Miley}, {Rocca-Volmerange}, {R{\"o}ttgering}, {Stanford}, {Teplitz}, {van
  Breugel}, \& {Zirm}}]{seymour07}
{Seymour} N., {Stern} D., {De Breuck} C., {Vernet} J., {Rettura} A.,
  {Dickinson} M., {Dey} A., {Eisenhardt} P., {Fosbury} R., {Lacy} M.,
  {McCarthy} P., {Miley} G., {Rocca-Volmerange} B., {R{\"o}ttgering} H.,
  {Stanford} S.~A., {Teplitz} H., {van Breugel} W., {Zirm} A., 2007,
  {Astrophys. J. Supplement Series}, 171, 353

\bibitem[{{Szomoru}(2008)}]{szomoru08}
{Szomoru} A., 2008, in The role of VLBI in the Golden Age for Radio Astronomy

\bibitem[{{van Weeren} {et~al.}(2009){van Weeren}, {R{\"o}ttgering},
  {Br{\"u}ggen}, \& {Cohen}}]{weeren09}
{van Weeren} R.~J., {R{\"o}ttgering} H.~J.~A., {Br{\"u}ggen} M., {Cohen} A.,
  2009, {Astron. Astrophys.}, 508, 75

\bibitem[{{Venemans} {et~al.}(2007){Venemans}, {R{\"o}ttgering}, {Miley}, {van
  Breugel}, {de Breuck}, {Kurk}, {Pentericci}, {Stanford}, {Overzier}, {Croft},
  \& {Ford}}]{venemans07}
{Venemans} B.~P., {R{\"o}ttgering} H.~J.~A., {Miley} G.~K., {van Breugel}
  W.~J.~M., {de Breuck} C., {Kurk} J.~D., {Pentericci} L., {Stanford} S.~A.,
  {Overzier} R.~A., {Croft} S., {Ford} H., 2007, {Astron. Astrophys.}, 461, 823

\end{thebibliography}

\label{lastpage}
\end{document}